\documentclass[preprint,12pt]{elsarticle}
\usepackage{graphicx} 
\usepackage{xparse}
\usepackage{amssymb, amsmath}
\usepackage{tikz}
\usepackage{mathtools}
\usepackage{physics}
\usepackage[margin=35truemm]{geometry}
\usepackage{stmaryrd}
\usepackage{version}
\numberwithin{equation}{section}

\usepackage{titlesec}

\titleformat{\subsection}
  {\bfseries}   
  {\thesubsection}
  {1em}
  {}

\usepackage{amsthm}
\theoremstyle{remark}
\newtheorem{remark}{\bf Remark}

\def\nn{\nonumber}
\def\Z{\mathbb{Z}}
\def\ua{\alpha}
\def\da{\dot{\alpha}}
\def\ub{\beta}

\def\db{\dot{\beta}}
\def\g{\mathfrak{g}}

\usepackage[dvips]{hyperref}
\hypersetup{
	colorlinks=true,
	citecolor=blue,
	linkcolor=blue,
	urlcolor=blue,
}

\journal{arXiv}

\biboptions{sort&compress}

\title{On Generalized Statistics and Stability in $\mathbb{Z}_2^2$-Graded Supersymmetric Yang--Mills Theory}
\author{Ren Ito, Akio Nago, and Shou Tanigawa
	\\[15pt]
	 Department of Physics, Osaka Metropolitan University,
\\Sugimoto Campus, Osaka 558-8585, Japan}

\begin{document}
\begin{abstract}
In the standard formulation of relativistic quantum field theory, a $\mathbb{Z}_2$-graded structure is assumed to realize locality and the boson--fermion dichotomy. 
While $\mathbb{Z}_2^n$-graded extensions are known to be allowed at the level of symmetry, their realization in interacting quantum field theories remains unclear.

In this paper, we construct a classical minimal $\mathbb{Z}_2^2$-graded supersymmetric Yang--Mills theory. 
We derive the invariant action and show that all kinetic terms have the correct sign, indicating the absence of classical ghost-like instabilities. 
Moreover, the positivity of the Hamiltonian follows from the $\mathbb{Z}_2^2$-graded supersymmetry algebra.

As a result, we show that $\mathbb{Z}_2^2$-graded generalized statistics can be realized at the classical level in a stable interacting supersymmetric gauge theory.

\end{abstract}
\maketitle
\section{Introduction}
Relativistic quantum field theory (QFT) is built upon a set of structural principles, including Lorentz invariance, locality, and unitarity. 
Within this framework, the spin--statistics connection plays a central role, as it justifies the conventional boson--fermion dichotomy. 
However, its derivation relies on the standard implementation of locality in terms of commutation and anticommutation relations, which already presupposes a $\mathbb{Z}_2$-graded structure. 
In this sense, the usual formulation of QFT incorporates such a binary grading at a fundamental level. 
This raises a fundamental question: is this binary grading intrinsic to quantum field theory, or merely a particular realization of locality?

Recent developments suggest that such a structure is not unique at the level of symmetry. 
In particular, $\mathbb{Z}_2^n$-graded Lie (super)algebras, introduced by Rittenberg and Wyler~\cite{RW1,RW2}, provide a natural generalization of ordinary Lie superalgebras. Here, $\mathbb{Z}_2^n$ denotes the $n$-fold direct product of the abelian group $\mathbb{Z}_2$. 
In our previous work~\cite{RIAN}, we showed that $\mathbb{Z}_2^n$-graded extensions of supersymmetry~\cite{Bruce}, as well as graded internal symmetries, can consistently arise as symmetries of the $S$-matrix, thereby extending the scope of the Coleman--Mandula~\cite{CM} and Haag--\L opusza\'nski--Sohnius theorems~\cite{HLS}.

These results indicate that generalized grading structures are allowed at the level of symmetry. 
The question is therefore no longer whether such structures are possible, but whether they can be realized as models in interacting relativistic QFT. 
Such a realization is highly nontrivial, since it would require an extension of the spin–statistics framework, which in conventional QFT underlies the compatibility of locality and energy positivity.

To date, the realization of such structures in relativistic interacting QFT remains largely unexplored. 
In particular, it is not known whether generalized statistics can be consistently implemented in gauge theories without spoiling energy positivity. 
For instance, Vasiliev's construction of de Sitter supergravity with $\mathbb{Z}_2^2$-graded symmetry~\cite{Vas} provides a remarkable example in which such a generalized grading arises naturally. 
In this case, ghost fields appear, indicating that consistency is highly nontrivial in relativistic settings.

By contrast, lower-dimensional realizations of $\mathbb{Z}_2^n$-graded structures have been investigated extensively in classical and quantum mechanics~\cite{BruDup,AAD,AKTcl,AKTqu,DoiAi1,AiItoTa2} and superfield formulations of lower-dimensional relativistic field theories~\cite{brusigma,bruSG,AIKT,AiItoTa,DoiAi2}, where explicit constructions have shown no instabilities associated with wrong-sign kinetic terms or negative-energy states. Moreover, $\mathbb{Z}_2^2$-graded supersymmetry leads to generalized statistics that can produce observable differences from the conventional bosonic and fermionic cases in multiparticle systems~\cite{Topp,Topp2}.

In this paper, we construct a classical minimal $\mathbb{Z}_2^2$-graded supersymmetric Yang--Mills theory using a superfield formulation. Supersymmetric Yang--Mills theory provides one of the most constrained frameworks in QFT, combining locality, gauge invariance, and supersymmetry in a highly rigid structure. Since supersymmetry ensures the positivity of the Hamiltonian, it also offers a natural setting to examine whether generalized statistics can be compatible with stability. Within this framework, we explicitly derive the invariant action and analyze its structure. In particular, all kinetic terms appear with the standard sign, indicating the absence of classical ghost-like instabilities. Moreover, the positivity of the Hamiltonian follows from the underlying $\mathbb{Z}_2^2$-graded supersymmetry algebra and is compatible with the structure of the action, even in the presence of fields with generalized $\mathbb{Z}_2^2$-graded exchange properties.

As a result, we show that generalized statistics can be realized consistently at the classical level in a stable interacting supersymmetric gauge theory. 
This provides a concrete realization of a class of theories whose existence in relativistic quantum field theory had remained unclear, and offers a new perspective on the role of grading structures in the foundations of QFT.

The organization of this paper is as follows. 
In Section 2, we review the algebraic structure of the minimal $\mathbb{Z}_2^2$-graded supersymmetry. 
In Section 2.1, we discuss the $\mathbb{Z}_2^2$-graded super-Poincar\'e algebra, and in Section 2.2, we construct the massless irreducible representation and identify the associated vector multiplet. 
In Section 3, we formulate the $\mathbb{Z}_2^2$-graded supersymmetry algebra in superspace. 
In Section 4, we present the superfield realization of the vector multiplet. 
In Section 5, we construct the invariant classical action and discuss its properties. 
Finally, Section 6 is devoted to concluding remarks.

\section{The $\Z_2^2$-graded supersymmetric algebra}
In this section, we first recall the definition of the minimal $\Z_2^2$-Supersymmetry algebra ($\Z_2^2$-SUSY algebra)  and $\Z_2^2$-super-Poincar\'e algebra, which is based on \cite{RIAN, Bruce}. Then we discuss a minimal irreducible multiplet for the $\Z_2^2$-SUSY algebra.
\subsection{Definition}
Let $\g$ be a $\Z_2^2$-graded vector space (over $\mathbb{R}$ or $\mathbb{C}$), which is the direct sum of homogeneous vector spaces labeled by $\Z_2^2:=\Z_2\times\Z_2=\{(0,0),(1,1),(0,1),(1,0)\}$:
\begin{align}
 \g=\g_{00}\oplus\g_{11}\oplus\g_{01}\oplus\g_{10}.
\end{align}
An element in $\g_{\vec{a}}~(\vec{a}\in\Z_2^2)$ is said to have the $\vec{a}$-grading.\footnote{
We denote $\mathfrak{g}_{(i,j)} \equiv \mathfrak{g}_{ij}$ for brevity, 
where $(i,j)\in \mathbb{Z}_2^2$.
Throughout this paper, we adopt the same shorthand notation for all generators and variables; for instance,
\[
Q^{(i,j)} \equiv Q^{ij}, 
\qquad 
\psi_{(i,j)} \equiv \psi_{ij},
\]
and similarly for other $\mathbb{Z}_2^2$-graded quantities.
}
Following the standard definition of $\mathbb{Z}_2^n$-graded Lie superalgebras~\cite{RW1,sch1},  a $\Z_2^2$-graded vector space equipped with the $\Z_2^2$-graded Lie bracket defined by
\begin{align}
 \llbracket A,B \rrbracket &\in \g_{\vec{a}+\vec{b}},
 ,\nn\\
 \llbracket A,B \rrbracket &= -(-1)^{\vec{a}\cdot\vec{b}}\llbracket B,A \rrbracket
 ,\nn\\
 \llbracket A,\llbracket B,C \rrbracket \rrbracket
 &=\llbracket \llbracket A,B \rrbracket ,C \rrbracket
 +(-1)^{\vec{a}\cdot\vec{b}}\llbracket B,\llbracket A,C \rrbracket \rrbracket
,
\end{align}
is called $\Z_2^2$-graded Lie superalgebra, where $A\in \g_{\vec{a}}$ and $\vec{a}\cdot\vec{b}$ is the standard scalar product of two dimensional vectors. We point out that the $\Z_2^2$-graded Lie bracket is realized as
\begin{align}
 \llbracket A, B \rrbracket =AB -(-1)^{\vec{a}\cdot\vec{b}} BA. \label{LieBracket}
\end{align}

$\Z_2^2$-graded Lie superalgebra can be regarded as an extended Lie algebra because the $\Z_2^2$-Lie  bracket \eqref{LieBracket} is a commutator (anticommutator) for $\vec{a}\cdot\vec{b}$ is even (odd). In fact, $\g_{00}$ is an ordinary Lie algebra, and $\g_{00}\oplus\g_{01}$ and $\g_{00}\oplus\g_{10}$ are ordinary Lie superalgebras.

A $\Z_2^2$-SUSY algebra is a particular example of a $\Z_2^2$-graded Lie superalgebra. In order to consider the simplest case, here we define the minimal $\Z_2^2$-SUSY algebra without any centers as 
\begin{align}
 \g=\text{lin. span.}\langle P_\mu, Q^{01}_{\alpha}, \bar Q^{01}_{\da}, Q^{10}_{\alpha}, \bar Q^{10}_{\da} \rangle,
\end{align}
where $P_\mu$ has $(0,0)$-grading, $Q^{01}_{\alpha}, \bar Q^{01}_{\da}$ have $(0,1)$-grading, and $Q^{10}_{\alpha}, \bar Q^{10}_{\da}$ have $(1,0)$-grading. 
The elements satisfy
\footnote{
We define $\eta_{\mu\nu}=\mathrm{diag}(-,+,+,+)$ and $(\sigma^\mu)_{\ua\da}=(\sigma^0,\sigma^1,\sigma^2,\sigma^3)_{\ua\da}$, where
\begin{align*}
 \sigma_0=\smqty(\pmat{0}),\qquad
\sigma_1= \smqty(\pmat{1})
,\qquad
\sigma_2= \smqty(\pmat{2})
,\qquad
\sigma_3= \smqty(\pmat{3}).
\end{align*}
Spinor indices are raised and lowered
using the antisymmetric tensors
$\varepsilon^{\alpha\beta}$ and $\varepsilon_{\alpha\beta}$,
with the conventions
\[
\psi^\alpha
=
\epsilon^{\alpha\beta}\psi_\beta,
\qquad
\psi_\alpha
=
\epsilon_{\alpha\beta}\psi^\beta,
\]
and similarly for dotted indices.
We take
\[
\varepsilon^{12} = -\varepsilon^{21} = 1,
\qquad
\varepsilon_{21} = -\varepsilon_{12} = 1.
\]
The tensors satisfy
$\epsilon^{\alpha\beta}\epsilon_{\beta\gamma}
= \delta^\alpha_\gamma$.
}
\begin{align}
 \{Q^{01}_{\ua}, \bar Q^{01}_{\da}\}&=\{Q^{10}_{\ua}, \bar Q^{10}_{\da}\}=2\sigma^\mu_{\ua\da}P_\mu
,\nn\\[5pt]
 \{Q^{01}_{\ua}, Q^{01}_{\ub}\}&=\{Q^{10}_{\ua}, Q^{10}_{\ub}\}=
[Q^{01}_{\ua}, Q^{10}_{\ub}]=[Q^{01}_{\ua}, \bar Q^{10}_{\da}]=0 \label{SUSYalg}
.
\end{align} 
Here, the ``minimal'' means that both of $\g_{00}\oplus\g_{01}$ and $\g_{00}\oplus\g_{10}$ are $\mathcal N=1$ SUSY algebra.

As seen in, there are two types of supercharges, with $(0,1)$- and $(1,0)$-gradings, in \eqref{SUSYalg}. In contrast to ordinary supersymmetry, supercharges with different $\mathbb{Z}_2^2$-gradings commute rather than anticommute, as dictated by the $\mathbb{Z}_2^2$-grading rule.
The $\mathbb{Z}_2^2$-graded structure marks a fundamental departure from ordinary supersymmetry. Theories endowed with symmetry generated by such an algebra naturally allow for states beyond the usual bosonic and fermionic sectors. 

We now introduce the $R$-symmetry generators $R^{00}$ and $R^{11}$ as automorphisms of the $\mathbb{Z}_2^2$-SUSY algebra, which satisfy
\begin{alignat}{4}
[P_\mu , R^{00}]&=0, \quad& [P_\mu, R^{11}]&=0, \quad& [R^{00}, R^{11}]&=0, \notag \\[2mm]
[Q^{01}_{\ua}, R^{00}]&=- Q^{01}_{\ua},\quad&  [\bar{Q}^{01}_{\da}, R^{00}]&= \bar Q^{01}_{\da},
\nn\\[2mm]
[Q^{10}_{\ua}, R^{00}]&=- Q^{10}_{\ua},\quad&  [\bar{Q}^{10}_{\da}, R^{00}]&= \bar Q^{10}_{\da},
\nn\\[2mm]
\{ Q^{01}_{\ua}, R^{11} \}&= Q^{10}_{\ua},\quad& \{ \bar{Q}^{01}_{\da}, R^{11} \}&= \bar Q^{10}_{\da}, \notag \\[2mm]
\{ Q^{10}_{\ua}, R^{11} \}&=Q^{01}_{\ua},\quad& \{ \bar{Q}^{10}_{\da}, R^{11} \}&=\bar Q^{01}_{\da}. \label{Ralg}
 \end{alignat}

The $\mathbb{Z}_2^2$-graded SUSY considered here possesses the same number of supercharges as ordinary $\mathcal{N}=2$ SUSY. 
While the R-symmetry algebra of conventional $\mathcal{N}=2$ SUSY is $\mathfrak{u}(1)\oplus\mathfrak{su}(2)$, in the present case the corresponding symmetry is instead generated by the $\Z_2^2$-graded elements $R^{00}$ and $R^{11}$. 
It should be emphasized that the $R^{11}$-action mixes the two $\mathbb{Z}_2^2$-supercharges, but it is not the continuous $SU(2)_R$ R-symmetry  of ordinary $\mathcal{N}=2$ SUSY.

Finally, we define the minimal $\Z_2^2$-super-Poinca\'e algebra. The ordinary Poincar\'e algebra satisfies the commutation relations
\begin{align}
 [P_\mu, M_{\nu\rho}]=&
-i(\eta_{\mu\nu} P_\rho   -\eta_{\rho\mu} P_\nu)
,\nn\\
[M_{\mu\nu},M_{\rho\sigma}]=&
i\left(\eta_{\mu\rho}M_{\nu\sigma}   -\eta_{\mu\sigma}M_{\nu\rho}   -\eta_{\nu\rho}M_{\mu\sigma}   +\eta_{\nu\sigma}M_{\mu\rho}\right).\label{Lor_alg}
\end{align}
To obtain the $\Z_2^2$-super-Poincar\'e algebra, we have to consider the relation between the supercharges and $M_{\mu\nu}$. From the Jacobi identities, we obtain these relations
\begin{align}
 [Q^{01}_{\ua},M_{\mu\nu}]&=i(\sigma_{\mu\nu})_\ua{^\ub} Q^{01}_\ub
 ,\quad
 [\bar Q^{01}_{\da}, M_{\mu\nu}]=-i\bar Q^{01}_{\db} (\bar \sigma_{\mu\nu})^{\db}{_{\da}}
,\nn\\
 [Q^{10}_{\ua},M_{\mu\nu}]&=i(\sigma_{\mu\nu})_\ua{^\ub} Q^{10}_\ub
 ,\quad
 [\bar Q^{10}_{\da}, M_{\mu\nu}]=-i\bar Q^{10}_{\db} (\bar \sigma_{\mu\nu})^{\db}{_{\da}}
,
 \label{Poin}
\end{align}
where $\left( \bar \sigma^\mu\right)^{\da\ua}:= \epsilon^{\da\db}\epsilon^{\ua\ub} \sigma^{\mu}_{\ub\db}$, and
\[
 (\sigma_{\mu\nu})_\ua{^\ub}:=\frac14\left(\sigma_{\mu\ua\da}\bar \sigma_{\nu}^{\da\ub}   -\sigma_{\nu\ua\da}\bar \sigma_{\mu}^{\da\ub}\right)
,\quad
(\bar \sigma_{\mu\nu})^{\da}{_{\db}}:=\frac14\left(\bar \sigma_{\mu}^{\da\ua}\sigma_{\nu\ua\db}   -\bar \sigma_{\nu}^{\da\ua}\sigma_{\mu\ua\db}\right).
\]
The $\Z_2^2$-graded Lie superalgebra, which satisfies the algebraic relations \eqref{SUSYalg}, \eqref{Lor_alg}, and \eqref{Poin}, is called the minimal $\Z_2^2$-super-Poinca\'e algebra. For the $R$-symmetry generators, one can see that 
\[
 [R^{00},M_{\mu\nu}]=[R^{11},M_{\mu\nu}]=0
\]
from Jacobi identities. This means that the $R$-symmetry is exactly an internal symmetry.

\subsection{Irreducible representations of the $\mathbb{Z}_2^2$-graded SUSY algebra}\label{sec.2.2}

In this section, we consider the massless irreducible representation. To make this construction more precise, we follow the standard Wigner method~\cite{S.W} and fix the momentum.
We first change the basis of the $\Z_2^2$-super-Poinca\'e algebra to
\[
  K_i:=M_{0i}, \quad J_{k}=\frac12\varepsilon_{kij}M_{ij},
\]
where $\varepsilon_{ijk}$ is the Levi-Civita tensor, defined $\varepsilon_{123}=1$. In this basis, the algebraic relations \eqref{Lor_alg} and \eqref{Poin} become
\begin{alignat}{4}
[J_i, J_j] &= i \epsilon_{ijk} J_k, \quad&
[J_i, K_j] &= i \epsilon_{ijk} K_k, \quad&
[K_i, K_j] &= - i \epsilon_{ijk} J_k, \nn\\
[J_i, P_j] &= i \epsilon_{ijk} P_k, \quad&
[K_i, P_0] &= -i P_i, \quad&
[K_i, P_j] &= -i \delta_{ij} P_0,
\end{alignat}
and
\begin{alignat}{3}
[Q^{01}_\alpha, J_i] \;&=\; \frac12 (\sigma^i)_\alpha{}^{\beta}\, Q^{01}_\beta,
&\qquad
[Q^{01}_\alpha, K_i] \;&=\; -\frac{i}{2} (\sigma^i)_\alpha{}^{\beta}\, Q^{01}_\beta,\nn\\
[\bar Q^{01}_{\dot\alpha}, J_i] \;&=\; -\frac12 \,\bar Q^{01}_{\dot\beta}\,(\sigma^i)^{\dot\beta}{}_{\dot\alpha},
&\qquad
[\bar Q^{01}_{\dot\alpha}, K_i] \;&=\; -\frac{i}{2}\,\bar Q^{01}_{\dot\beta}\,(\sigma^i)^{\dot\beta}{}_{\dot\alpha},
\nn\\
[Q^{10}_\alpha, J_i] \;&=\; \frac12 (\sigma^i)_\alpha{}^{\beta}\, Q^{10}_\beta,
&\qquad
[Q^{10}_\alpha, K_i] \;&=\; -\frac{i}{2} (\sigma^i)_\alpha{}^{\beta}\, Q^{10}_\beta,\nn\\
[\bar Q^{10}_{\dot\alpha}, J_i] \;&=\; -\frac12 \,\bar Q^{10}_{\dot\beta}\,(\sigma^i)^{\dot\beta}{}_{\dot\alpha},
&\qquad
[\bar Q^{10}_{\dot\alpha}, K_i] \;&=\; -\frac{i}{2}\,\bar Q^{10}_{\dot\beta}\,(\sigma^i)^{\dot\beta}{}_{\dot\alpha} \label{rel_QJ}.
\end{alignat}

For the massless case, we choose the representative momentum
\[
  P_\mu\ket{p}=p_\mu\ket{p},\quad p_\mu=(-E,0,0,E).
\]
Now $\ket{p}$ can have any $\Z_2^2$-grading. We will fix this grading later for physical interpretation.
In this representation, $|p\rangle$ transforms under a representation of the little group $ISO(2)$, which satisfies 
\[
 [J_3, L_\pm]=\pm i L_\mp,\quad [L_+,L_-]=0,
\]
where $L_+:=K_1+J_2$ and $L_-:=K_2-J_1$. That is because 
\[
 [P_\mu, J_3]\ket{p}=[P_\mu, L_+]\ket{p}=[P_\mu, L_-]\ket{p}=0.
\]
This means that $|p\rangle$ furnishes a representation of the little group $ISO(2)$. Thus, we assume 
\begin{align}
 J_3\ket{h}=h\ket{h},\quad L_\pm \ket{h}=0, \label{irrep_PA}
\end{align}
where we denote $\ket{p,h}\equiv \ket{h}$.
Then the $\ket{h}$ only depends on the eigenvalue of $h$, called helicity.

Moreover, as seen in the previous section, the $R$-symmetry is an internal symmetry, which means that $\ket{h}$ also furnishes a representation of the $R$-symmetry group.
Here, two possibilities arise for the action of the $R$-symmetry generators on the state $|h\rangle$.
First, $|h\rangle$ may be chosen as a simultaneous eigenstate of
$R^{00}$ and $R^{11}$:
\begin{align}
 R^{00}\ket{h,r} &= r\,\ket{h,r}, \qquad
 R^{11}\ket{h,r} = 0.
\end{align}
Alternatively, the action of $R^{11}$ may generate an independent state
$|\lambda\rangle'$,
\begin{align}
 R^{00}\ket{h,r,s} &= r\,\ket{h,r,s}, \nn\\
 \ket{h,r,s}' :&= R^{11}\ket{h,r,s}, \qquad
 R^{11}\ket{h,r,s}' = s\,\ket{h,r,s},
\end{align}
so that the pair $\{|h,r,s\rangle,|h,r,s\rangle'\}$
forms a two-dimensional representation under the action of $R^{11}$.
In the following, we restrict ourselves to the first (simpler) case,
in which the state $|h,r\rangle$ is an eigenstate
of the $R$-symmetry generators.

We now consider the action of the supercharges.
Substituting $p_\mu=(-E,0,0,E)$ into the $\mathbb{Z}_2^2$-SUSY algebra \eqref{SUSYalg},
we obtain
\begin{align}
 \{Q^{01}_{1}, \bar Q^{01}_{\dot 1}\}
 &= \{Q^{10}_{1}, \bar Q^{10}_{\dot 1}\}
 = 4E,
 \nn\\
 \{Q^{01}_{2}, \bar Q^{01}_{\dot 2}\}
 &= \{Q^{10}_{2}, \bar Q^{10}_{\dot 2}\}
 = 0.
\end{align}
Thus, only the first components $Q^{01}_1$ and $Q^{10}_1$
act non-trivially and we may therefore impose the condition
\[
 Q^{01}_{2}\ket{h,r}= \bar Q^{01}_{\dot 2}\ket{h,r}=0.
\]
Moreover, using the relation \eqref{rel_QJ}, one can find that $Q^{01}_{\ua}$ and $Q^{10}_{\ua}$ lower the helicity by $\tfrac{1}{2}$, while $\bar Q^{01}_{\da}$ and $\bar Q^{10}_{\da}$ raise the helicity by $\tfrac{1}{2}$. Thus, we can define the highest helicity state as
\begin{equation}
 \bar Q^{01}_{\dot 1} \ket{h,r}=\bar Q^{10}_{\dot 1} \ket{h,r}=0,
\end{equation}
so that the state $\ket{h,r}$ is annihilated by the conjugate supercharges.
The irreducible multiplet is therefore generated by acting
with the supercharges $Q^{01}_{\ua}$ and $Q^{10}_{\ua}$ on the state $|h,r\rangle$.

Under this assumption, using the relations \eqref{rel_QJ} and \eqref{Ralg} the massless irreducible multiplet is obtained as the helicity and the $R^{00}$-charge ladder
\begin{align}
&\bigl\{
|h,r\rangle,\;
Q^{01}_{1}|h,r\rangle,\;
Q^{10}_{1}|h,r\rangle,\;
Q^{01}_{1}Q^{10}_{1}|h,r\rangle
\bigr\}
\nn\\
 =:&\big\{   \ket{h,r},\,   \ket{h-\tfrac12 ,r+1}_1,\,   \ket{h-\tfrac12,r+1}_2,\,   \ket{h-1,r+2}   \big\}.\label{VM}
\end{align}
Their helicities are
$h$, $h-\tfrac12$,
$h-\tfrac12$, and $h-1$, and $R^{00}$-charges are $r$, $r+1$, $r+1$, $r+2$, respectively.
Since $Q^{01}_1$ and $Q^{10}_1$ commute and satisfy
$(Q^{01}_1)^2=(Q^{10}_1)^2=0$,
no further independent states can be generated.
These four states therefore exhaust the minimal irreducible representation.

For applications to gauge theory, we take the highest helicity
to be $h=+1$ and choose the reference state $|+1\rangle$
to have $(0,0)$-grading.
Under this choice, the helicity ladder \eqref{VM} reads
\begin{equation}
\bigl\{
|+1,r\rangle_{00},\;
|+\tfrac12,r+1\rangle_{01},\;
|+\tfrac12,r+1\rangle_{10},\;
|0,r+2\rangle_{11}
\bigr\}
,
\end{equation}
with helicities $+1$, $+\tfrac12$, $+\tfrac12$, and $0$, respectively.

After CPT completion, the multiplet contains states of helicities
$\pm1$, $\pm\tfrac12$, and $0$.
The two helicity states $\pm1$ correspond to the two physical
polarizations of a single massless gauge field $A_\mu$,
rather than to distinct fields.
Thus, the field content and their $\mathbb{Z}_2^2$-gradings are summarized as follows:
\begin{equation}
\begin{array}{c|c|c|c}
\text{Field} & \text{Helicity} & R^{00}\text{-charge} & \mathbb{Z}_2^2\text{ grading} \\ \hline
A_\mu & \pm1 & r & (0,0) \\
\lambda_{01} & \pm\tfrac12 & r+1 & (0,1) \\
\psi_{10} & \pm\tfrac12 & r+1 & (1,0) \\
\varphi_{11} & 0 & r+2 & (1,1)
\end{array}\label{multiplet}
\end{equation}
We emphasize that the $\mathbb{Z}_2^2$-grading is independent of the helicity and $R$-charge assignment. In particular, states with identical helicity may have different $\mathbb{Z}_2^2$-gradings.

Now let us comment on the action of the $R$-symmetry on the fields.
From the table \eqref{multiplet}, one can read off the field transformations by $R^{00}$
\begin{alignat}{2}
 \delta_{R^{00}}A_\mu &= i\gamma_{00}\cdot r A_\mu
,\quad& 
 \delta_{R^{00}}(\lambda_{01})_{\ua} &= i\gamma_{00}\cdot (r+1)(\lambda_{01})_{\ua}
,\nn\\
 \delta_{R^{00}}(\psi_{10})_{\ua} &= i\gamma_{00}\cdot (r+1)(\psi_{10})_{\ua}
,\quad& 
 \delta_{R^{00}}\varphi_{11} &= i\gamma_{00}\cdot (r+2) \varphi_{11},
\label{R00sym}
\end{alignat}
where $\delta_{R^{00}}:=i\gamma_{00}R^{00}$ and $\gamma_{00}$ is the transformation parameter.

Then, from the algebraic relations \eqref{Ralg}, we obtain the action of the $R^{11}$
\begin{alignat}{2}
 R^{11}\ket{-\tfrac12,r+1}_{01}&=\ket{-\tfrac12,r+1}_{10}
,\quad&
 R^{11}\ket{-\tfrac12,r+1}_{10}&=\ket{-\tfrac12,r+1}_{01}
,\nn\\
  R^{11}\ket{-1,r}_{00}&=0
,\quad&
R^{11}\ket{0,r+2}_{11}&=0.
\label{R11sym}
\end{alignat}
From these relations, one can read off the corresponding field transformations by $R^{11}$ 
\begin{alignat}{2}
 \delta_{R^{11}}A_\mu &= 0
,\quad& 
 \delta_{R^{11}}(\lambda_{01})_{\ua} &= i\gamma_{11}(\psi_{10})_{\ua}
,\nn\\
 \delta_{R^{11}}(\psi_{10})_{\ua} &= i\gamma_{11}(\lambda_{01})_{\ua}
,\quad& 
 \delta_{R^{11}}\varphi_{11} &= 0
\end{alignat}
where $\delta_{R^{11}}:=i\gamma_{11}R^{11}$ and $\gamma_{11}$ is the transformation parameter.

Thus, the transformation $\delta_{R^{00}}$ can be interpreted as a $U(1)$ $R$-symmetry. 
The pair $\{(\lambda_{01})_\ua, (\psi_{10})_\ua\}$ forms a doublet representation  under the action of $R^{11}$. This mixing differs from the usual $SU(2)_R$ symmetry of ordinary 
$\mathcal{N}=2$ SUSY. 
Here, it arises from the $\mathbb{Z}_2^2$-graded automorphism generated by $R^{11}$.

Finally, we refer to this representation as the 
minimal $\mathbb{Z}_2^2$-vector multiplet.
It consists of a massless gauge field $A_\mu$, two fermionic partners with the $(0,1)$- and $(1,0)$-grading $\{(\lambda_{01})_\ua, (\psi_{10})_{\ua}\}$, and a $(1,1)$-graded scalar field $\varphi_{11}$.
The fermionic fields are mixed under the action of $R^{11}$.

\section{$\Z_2^2$-graded superspace}

In this section, we construct a superspace realization of the  $\mathbb{Z}_2^2$-SUSY algebra (\Ref{SUSYalg}). 
Our construction follows the standard superspace formalism~\cite{S.W, WessBagger}, 
appropriately generalized to accommodate the $\mathbb{Z}_2^2$-grading structure. 
Lower-dimensional $\mathbb{Z}_2^2$-graded superspace constructions were previously discussed in~\cite{brusigma, bruSG,AIKT,AiItoTa,DoiAi2}, 
but a $(1+3)$-dimensional realization has not been constructed so far.

We begin by introducing a $\Z_2^2$-superspace
$\mathfrak{S}$ defined by
\begin{align}
    \mathfrak{S} = (x^\mu, \xi^{\ua}, \bar \xi^{\da}, \eta^{\ua}, \bar \eta^{\da}),
\end{align}
where $\xi$ has $(0,1)$-grading and $\eta$ has $(1,0)$-grading. The (anti)commutation relations are determined
by the $\Z_2^2$-grading:
\begin{align}
 \{\xi^\alpha, \xi^{\beta}\}&=\{\xi^\alpha, \bar\xi^{\dot\beta}\}=
 \{\eta^\alpha, \eta^{\beta}\}=\{\eta^\alpha, \bar\eta^{\dot\beta}\}=0
,\nn\\
[\xi^\alpha, \eta^{\beta}]&=[\xi^\alpha, \bar\eta^{\dot\beta}]=0.\label{coordinatealg}
\end{align}
We adopt the following complex conjugation conventions:
\[
    (x^\mu)^*=x^\mu,\quad (\xi^{\ua})^*=\bar \xi^{\da},\quad (\eta^{\ua})^*=\bar \eta^{\da},
\]
where the involution ``$*$'' is defined by
\begin{align}
    \left(s_{\vec{a}}^* \right)^* = s_{\vec{a}},\quad (s_{\vec{a}}s_{\vec{b}})^*= s_{\vec{b}}^* s_{\vec{a}}^*,\quad s_{\vec{a}}\in \mathfrak{S}. \label{eq:cc}
\end{align}

We next introduce derivatives with respect to each coordinate:
\[
    \partial_\mu:=\frac{\partial}{\partial x^\mu},\quad
    \partial_{\theta^{\ua}} \theta^{\ub} = \delta_\ua^\ub,\quad \partial_{\bar \theta^{\da}} \bar\theta^{\db} = \delta_{\da}^{\db},
    \quad \theta^{\ua}\in\{\xi^{\ua}, \eta^{\ua}\}.
\]
The set of coordinates and derivatives forms a $\Z_2^2$-graded Lie superalgebra
with the relations \eqref{coordinatealg} and
\begin{align}
    [\partial_\mu, x^\nu]=\delta_{\mu}^\nu
    ,\quad
    \{\partial_{{\xi}^{\ua}},\xi^{\ub}\}
    =\{\partial_{\eta^{\ua}},\eta^{\ub}\}
    =\delta_{\ua}^{\ub}
    ,\quad
    \{\partial_{\bar \xi^{\da}},\bar \xi^{\db}\}
    =\{\partial_{\bar \eta^{\da}},\bar \eta^{\db}\}
    =\delta_{\da}^{\db}
    .
\end{align}
All other (anti)commutators vanish. The complex conjugation acts as an automorphism of this algebra. Consequently, the derivatives satisfy
\begin{align}
    \left( \partial_\mu \right)^*=-\partial_\mu,\quad
    \left( \partial_{\xi^{\ua}} \right)^*=\partial_{\bar \xi^{\da}},\quad
    \left( \partial_{\eta^{\ua}} \right)^*=\partial_{\bar \eta^{\da}}.
\end{align}

We now introduce differential operators on superspace which realize the $\Z_2^2$-SUSY algebra.
The corresponding $\Z_2^2$-SUSY transformations on the superspace coordinates are defined by
\[
    \delta \mathfrak{S}:= \mathfrak{S}'-\mathfrak{S}
\]
with
\begin{align} 
 \delta x^\mu
 & :=\varepsilon_{00}^\mu 
 -i(\varepsilon_{01}\sigma^\mu \bar \xi)+i(\xi\sigma^\mu \bar \varepsilon_{01})
 -i(\varepsilon_{10}\sigma^\mu \bar \eta)+i(\eta\sigma^\mu \bar \varepsilon_{10})
 ,\nn\\
 \delta \xi^{\ua} &:=\varepsilon_{01}^{\ua}
 ,\quad
 \delta \bar \xi^{\da} :=\bar \varepsilon_{01}^{\da}
 ,\quad
 \delta \eta^{\ua} :=\varepsilon_{10}^{\ua} 
 ,\quad
 \delta \bar \eta^{\ua} :=\bar\varepsilon_{10}^{\ua}.
\end{align}
Then we define the differential generators by
\begin{align} \label{tf:susy}
    \delta_{00} \mathfrak{S}
      :&= \big[i\varepsilon_{00}^\mu P_\mu,~ \mathfrak{S}\big]
    ,\nn\\
    \delta_{01} \mathfrak{S}
      :&= \left[\varepsilon_{01}Q^{01} - \bar Q^{01} \bar \varepsilon_{01},~ \mathfrak{S}\right]
    ,\nn\\
    \delta_{10} \mathfrak{S}
    :&= \left[\varepsilon_{10}Q^{10} - \bar Q^{10} \bar \varepsilon_{10},~ \mathfrak{S}\right],
\end{align}
where $\varepsilon_{\vec{a}}$ denotes the transformation parameter of  $\vec{a}$-grading. Spinor contractions are defined with the $\Z_2^2$-graded sign rule: 
\begin{align*}
 \psi_{\vec{a}}\lambda_{\vec{b}}&=(\psi_{\vec{a}})^\alpha(\lambda_{\vec{b}})_{\alpha}=-(-1)^{\vec{a}\cdot\vec{b}}(\lambda_{\vec{b}})^\alpha(\psi_{\vec{b}})_{\alpha}=-(-1)^{\vec{a}\cdot\vec{b}}\lambda_{\vec{b}}\psi_{\vec{a}}
,\\
\bar \psi_{\vec{a}}\bar \lambda_{\vec{b}}&=(\bar \psi_{\vec{a}})_{\da}(\bar \lambda_{\vec{b}})^{\da}=-(-1)^{\vec{a}\cdot\vec{b}}(\bar \lambda_{\vec{b}})_{\da}(\bar \psi_{\vec{b}})^{\da}=-(-1)^{\vec{a}\cdot\vec{b}}\bar \lambda_{\vec{b}}\bar \psi_{\vec{a}}
.
\end{align*}
By definition, one can read the explicit forms of the generators:
\begin{align}\label{op:susy}
    P_\mu
    :=-i\partial_\mu
    ,\quad
    Q_{\ua}^{01}
     & :=\partial_{\xi^\ua}-i\left(\sigma^\mu \bar \xi\right)_{\ua} \partial_\mu
    , \quad
    \bar Q_{\da}^{01}
    :=\partial_{\bar{\xi}^{\da}}-i\left(\xi \sigma^\mu\right)_{\da} \partial_\mu,\nn
    \\
    Q_{\ua}^{10}
     & :=\partial_{\eta^\ua}-i\left(\sigma^\mu \bar \eta\right)_{\ua} \partial_\mu
    , \quad
    \bar Q_{\da}^{10}
    :=\partial_{\bar{\eta}^{\da}}-i\left(\eta \sigma^\mu\right)_{\da} \partial_\mu.
\end{align}
By direct computation, one verifies that the generators \eqref{op:susy} satisfy the algebraic relations of the $\Z_2^2$-SUSY algebra \eqref{SUSYalg}.
Hence, $\delta_{01}\mathfrak{S}$ and $\delta_{10}\mathfrak{S}$ realize the $\Z_2^2$-SUSY on superspace.

Finally, we introduce covariant derivatives which (anti)commute with the supercharges:
\begin{alignat}{2}\label{cov.der}
    D_{\ua}^{01}
     & =\partial_{\xi^\ua}+i\left(\sigma^\mu \bar \xi\right)_{\ua} \partial_\mu
    , \quad&
    \bar D_{\da}^{01}
    &=\partial_{\bar{\xi}^{\da}}+i\left(\xi \sigma^\mu\right)_{\da} \partial_\mu,\nn
    \\
    D_{\ua}^{10}
     & =\partial_{\eta^\ua}+i\left(\sigma^\mu \bar \eta\right)_{\ua} \partial_\mu
    , \quad&
    \bar D_{\da}^{10}
    &=\partial_{\bar{\eta}^{\da}}+i\left(\eta \sigma^\mu\right)_{\da} \partial_\mu. 
\end{alignat}
Then they satisfy
\begin{align}
  \{D^{01}_{\ua}, \bar D^{01}_{\da}\}&=\{D^{10}_{\ua}, \bar D^{10}_{\da}\}=-2\sigma^\mu_{\ua\da}P_\mu
,\nn\\[5pt]
 \{D^{01}_{\ua}, D^{01}_{\ub}\}&=\{D^{10}_{\ua}, D^{10}_{\ub}\}=
[D^{01}_{\ua}, D^{10}_{\ub}]=[D^{01}_{\ua}, \bar D^{10}_{\da}]=0 
,
\end{align}
and
\begin{align}
  \{Q^{01}_{\ua}, \bar D^{01}_{\da}\}&=\{Q^{10}_{\ua}, \bar D^{10}_{\da}\}=
 \{Q^{01}_{\ua}, D^{01}_{\ub}\}=\{Q^{10}_{\ua}, D^{10}_{\ub}\}=0
,\nn\\
[Q^{01}_{\ua}, D^{10}_{\ub}]&=[Q^{01}_{\ua}, \bar D^{10}_{\da}]=0 
.
\end{align}
This completes the superspace realization of the $\Z_2^2$-SUSY algebra, which will be used in the construction of superfields in the next section.

\section{$\Z_2^2$-Superfield formulation}
In this section we construct superfields on superspace in the previous section.
Superfields are $\Z_2^2$-graded functions on $\mathfrak S$, which are invariant under the $\Z_2^2$-SUSY transformation \eqref{tf:susy}, i.e. $F'(\mathfrak S')=F(\mathfrak S)$.
By definition, we can describe the transformation of the superfield as
\begin{align}
    \delta F(\mathfrak{S})
     & :=F'(\mathfrak{S})-F(\mathfrak{S})\nn
    \\
     & =-(\delta_{00}+\delta_{01}+\delta_{10})F(\mathfrak{S}). \label{tf:field}
\end{align}
We first analyze the $(0,1)$-sector, which parallels the ordinary $\mathcal N=1$ SUSY case, and then extend the construction to the full version of the minimal $\Z_2^2$-vector multiplet.
\subsection{The chiral superfield}
We follow the standard $\mathcal N=1$ superfield construction~\cite{WessBagger} appropriately generalized to the $\mathbb{Z}_2^2$-graded setting.
As discussed in Section \ref{sec.2.2}, the scalar component of the vector multiplet has the $(1,1)$-grading. In this sense, here we consider the $(1,1)$-graded chiral superfield $\Phi_{11}$ on $(x^\mu, \xi^{\ua}, \bar \xi^{\da})$, which satisfies $\bar D^{01}_{\dot\alpha}\Phi_{11}=0$.
For simplicity, we introduce the coordinate $y=x+i(\xi \sigma^\mu \bar \xi)$. 
In terms of this coordinate, the $(0,1)$-SUSY generators and the covariant derivatives take the form
\begin{align}
    Q_{\ua}^{01}
     & =\partial_{\xi^\ua}
    , \quad
    \bar Q_{\da}^{01}
    =\partial_{\bar{\xi}^{\da}}-2i\left(\xi \sigma^\mu\right)_{\da} \partial_\mu,
\label{op:susy_y}\\
    D_{\ua}^{01}
     & =\partial_{\xi^\ua}+2i\left(\sigma^\mu \bar \xi\right)_{\ua} \partial_\mu
    , \quad
    \bar D_{\da}^{01}
    =\partial_{\bar{\xi}^{\da}}
.\label{cov.der_y}
\end{align}
With these expressions, the chiral superfield can be written as
\begin{align}
    \Phi_{11}(y,\xi)=\varphi_{11}(y)+\sqrt{2}\xi\psi_{10}(y)-\xi\xi F_{11}(y),
\end{align}
which indeed satisfies $\bar D^{01}_{\da} \Phi_{11}=0 $.
From this expansion, one identifies the component fields of the chiral multiplet. The field $\varphi_{11}$ is a complex scalar
with $(1,1)$-grading, $(\psi_{10})_{\alpha}$ is its $(1,0)$-graded fermionic partner, and $F_{11}$ is a complex auxiliary field. The auxiliary field $F_{11}$ does not propagate and becomes algebraically determined once the invariant action is specified.
Using \eqref{tf:field}, we obtain the transformations of components:
\begin{align}
    \delta_{01}\varphi_{11}
     & =\sqrt{2}\varepsilon_{01}\psi_{10},\nn
    \\
    \delta_{01}\left(\psi_{10}\right)_{\ua}
     & =-\sqrt{2}\left(\varepsilon_{01}\right)_{\ua}F_{11}
    +\sqrt{2}i\left(\sigma^{\mu} \bar \varepsilon_{01} \right)_{\ua}\partial_\mu\varphi_{11},\nn
    \\
    \delta_{01}F_{11}
     & = -\sqrt{2}i \left(\partial_\mu \psi_{10} \sigma^\mu \bar \varepsilon_{01}\right).\label{Phi01SUSY1}
\end{align}

On the other hand, by introducing $\bar y=x-i(\xi \sigma^\mu \bar \xi)$, we can easily describe the anti-chiral superfield which satisfies $D^{01}_{\ua} \bar \Phi_{11}=0$ as
\begin{align}
    \bar \Phi_{11}(\bar y,\bar \xi)
    =\bar \varphi_{11}(\bar y)+\sqrt{2}\bar \psi_{10}(\bar y) \bar \xi- \bar F_{11}(\bar y) \bar \xi \bar \xi,
\end{align}
with
\begin{align}\label{cov.der_bar_y}
    D_{\ua}^{01}
     & =\partial_{\xi^\ua}
    , \quad
    \bar D_{\da}^{01}
    =\partial_{\bar{\xi}^{\da}}+2i\left(\xi \sigma^\mu\right)_{\da} \partial_\mu
    .
\end{align}
Then the SUSY generators are rewritten as
\begin{align}\label{op:susy_bar_y}
    Q_{\ua}^{01}
     & =\partial_{\xi^\ua}-2i\left(\sigma^\mu \bar \xi\right)_{\ua} \partial_\mu
    , \quad
    \bar Q_{\da}^{01}
    =\partial_{\bar{\xi}^{\da}}
.
\end{align}
Using \eqref{tf:field}, we also obtain the transformations of components:
\begin{align}
    \delta_{01}\bar \varphi_{11}
     & =\sqrt{2}\bar \psi_{10} \bar \varepsilon_{01},\nn
    \\
    \delta_{01}\left(\bar \psi_{10}\right)_{\da}
     & =-\sqrt{2}\bar F_{11} \left(\bar \varepsilon_{01}\right)_{\da}
    -\sqrt{2}i \partial_\mu\bar \varphi_{11} \left( \varepsilon_{01} \sigma^{\mu} \right)_{\da},\nn
    \\
    \delta_{01} \bar F_{11}
     & = \sqrt{2}i \left(\varepsilon_{01} \sigma^\mu \partial_\mu \bar \psi_{10} \right).\label{barPhi01SUSY}
\end{align}
The transformation \eqref{barPhi01SUSY} is the complex conjugate of \eqref{Phi01SUSY1}.

\subsection{The vector superfield}
As discussed in Section \ref{sec.2.2}, the gauge field of the vector multiplet has the $(0,0)$-grading. In this section, we introduce a real (0,0)-graded vector superfield $V$ on $(x^\mu, \xi^{\ua}, \bar \xi^{\da})$, which is real, $\bar V(x,\xi,\bar \xi)=V(x,\xi,\bar \xi)$. This follows the standard $\mathcal N=1$ vector superfield construction~\cite{WessBagger}.
It takes the form
\begin{align}
    V (x,\xi,\bar{\xi})
     & = C^{} +i \xi \chi_{01} -i \bar \xi \bar \chi_{01}
    + \frac{i}{2} \xi\xi M -\frac{i}{2} \bar \xi \bar \xi \bar M -\left(\xi \sigma^\mu \bar \xi \right)A_\mu
    \nn\\
     & + i\xi\xi \bar \xi \left( \bar \lambda_{01} + \frac{i}{2}\bar \sigma^\mu \partial_\mu \chi_{01} \right)
    - i\bar \xi \bar \xi \xi \left( \lambda_{01} + \frac{i}{2}\sigma^\mu \partial_\mu \bar \chi_{01} \right)
    \nn\\
    &+\frac{1}{2}\xi\xi \bar\xi\bar\xi \left(D+\frac{1}{2} \Box C \right).
\end{align}
For a non-Abelian $SU(N)$ gauge theory, the vector superfield takes values in the Lie algebra $\mathfrak{su}(N)$, i.e. $V=V^a T^a$. The gauge transformation is given by
\begin{align}
e^{2gV} \to   e^{-2i g\bar\Lambda} e^{2gV} e^{2i g\Lambda}
,\qquad
e^{-2gV} \to   e^{-2i g\Lambda} e^{-2gV} e^{2 ig\bar\Lambda}
,
\end{align} 
where $\Lambda (x, \xi,\bar\xi)$ is a $(0,0)$-graded chiral superfield, which satisfies $\bar D^{01}_{\da}\Lambda=0$ with \eqref{cov.der}.
From this transformation, we obtain the corresponding transformation of $V$:
\begin{align}
V \to V+i(\Lambda-\bar\Lambda) +g\Bigl([\bar\Lambda,\Lambda] +i\,[V,\Lambda+\bar\Lambda]\Bigr) +O(g^2).
\label{eq:Vgauge_Og}
\end{align}
We now impose the Wess--Zumino gauge in analogy with the ordinary case~\cite{WessBagger}.
We define the $(0,0)$-graded chiral superfield by
\begin{align}
 \Lambda=\sum_{n=0}^{\infty} g^n\Lambda_n.
\end{align} 
Using this expression, the RHS of \eqref{eq:Vgauge_Og} becomes
\begin{align}
 V+i(\Lambda_0-\bar\Lambda_0)   +g\Bigl([\bar\Lambda_0,\Lambda_0] +i\,[V,\Lambda_0+\bar\Lambda_0] +i(\Lambda_1 -\bar\Lambda_1)\Bigr) +O(g^2). \label{Vgauge}
\end{align}
If we define 
\begin{align}
 \Lambda_0 (x,\xi,\bar\xi)
=&\frac{i}{2}C-\xi\chi_{01}-\frac12 \xi\xi M 
\nn\\
&-\frac12 \xi\sigma^\mu\bar\xi \partial_\mu C -\frac{i}{2}\xi\xi\bar\xi \bar\sigma^\mu \partial_\mu \chi_{01} +\frac{i}{8}\xi\xi\bar\xi\bar\xi\, \Box C,
\end{align}
then we obtain
\begin{align}
i(\Lambda_0-\bar\Lambda_0)
=&-\,C
- i\,\xi\chi_{01}
+ i\,\bar\xi\,\bar\chi_{01}
- \frac{i}{2}\,\xi\xi\,M
+ \frac{i}{2}\,\bar\xi\bar\xi\,\bar M
\nonumber\\
&
+\frac12\,\xi\xi\bar\xi\,\bar\sigma^\mu\partial_\mu\chi_{01}
-\frac12\,\bar\xi\bar\xi\xi\,\sigma^\mu\partial_\mu\bar\chi_{01}
-\frac14\,\xi\xi\bar\xi\bar\xi\,\Box C ,
\end{align}
and the terms at zeroth order in the coupling constant $g$ in \eqref{Vgauge} take
\begin{align}
 V+i(\Lambda_0-\bar\Lambda_0)
 = -\left(\xi \sigma^\mu \bar \xi \right)A_\mu\nn
 + i\xi\xi \bar \xi \bar \lambda_{01}
 - i\bar \xi \bar \xi \xi  \lambda_{01}
 +\frac{1}{2}\xi\xi \bar\xi\bar\xi D,
\end{align}
where $\Lambda_0(x,\xi,\bar\xi)$ satisfies $\bar D^{01}_{\da}\Lambda_0=0$ with \eqref{cov.der}. 
At each order in the coupling constant $g$, the corresponding terms of $g^n$ ($n\ge1$) can be removed by an appropriate choice of the higher-order correction $\Lambda_n$ in the expansion of the gauge parameter.
Therefore, we obtain a vector superfield in the Wess--Zumino gauge:
\begin{align}
    V_{\text{WZ}} (x,\xi,\bar{\xi})
      = -\left(\xi \sigma^\mu \bar \xi \right)A_\mu\nn
     + i\xi\xi \bar \xi \bar \lambda_{01}
    - i\bar \xi \bar \xi \xi  \lambda_{01}
    +\frac{1}{2}\xi\xi \bar\xi\bar\xi D.
\end{align}

However, the $(0,1)$-graded SUSY transformation does not preserve the Wess--Zumino gauge. Acting on $V_{\rm WZ}$, one finds
\begin{align*}
\delta_{01} V_{\rm WZ}
&= -\,\xi \sigma^\mu \bar{\varepsilon}_{01}\, A_\mu
   - A_\mu \,\varepsilon_{01} \sigma^\mu \bar\xi
   + i\, \xi\xi\, \bar{\lambda}_{01}\,\bar{\varepsilon}_{01}
   - i\, \bar{\xi}\bar{\xi}\, \varepsilon_{01} \lambda_{01}
\nonumber\\[4pt]
&\quad
   - (\xi \sigma^\mu \bar{\xi})
     \bigl(
        i\,\varepsilon_{01} \sigma_\mu \bar{\lambda}_{01}
        - i\,\lambda_{01} \sigma_\mu \bar{\varepsilon}_{01}
     \bigr)
   + \xi\xi \bar{\xi}
     \left(
        \bar{\varepsilon}_{01} D
        - \frac{i}{2}\,
          \bar{\sigma}^\mu \sigma^\nu \bar{\varepsilon}_{01}\,
          \partial_\nu A_\mu
     \right)
\nonumber\\[4pt]
&\quad
   + \bar{\xi}\bar{\xi}\xi
     \left(
        \varepsilon_{01} D
        + \frac{i}{2}\,
          \sigma^\mu \bar{\sigma}^\nu \varepsilon_{01}\,
          \partial_\nu A_\mu
     \right)
   - \frac{1}{2}\,\xi\xi \bar{\xi}\bar{\xi}
     \left(
        \varepsilon_{01} \sigma^\mu \partial_\mu \bar{\lambda}_{01}
        + \partial_\mu \lambda_{01} \sigma^\mu \bar{\varepsilon}_{01}
     \right)
,
\end{align*}
which generates $\bar\xi_{\dot\alpha}$- and $\bar\xi\bar\xi$-components.
Because such components are eliminated by the Wess--Zumino gauge condition, the gauge fixing is violated.
Therefore, we have to consider a compensating gauge transformation associated with the SUSY transformation, so that the gauge condition is preserved.
To this end, we define $\Lambda^{(1)}$ as
\begin{align}
\Lambda^{(1)}(x,\xi,\bar\xi)
&=- i(\xi\sigma^\mu\bar\varepsilon_{01})A_\mu
- \xi\xi\,\bar\lambda_{01}\bar\varepsilon_{01}
 +\frac12 \xi\xi\bar\xi\bar \sigma^\nu\sigma^\mu \bar \varepsilon_{01}\partial_\nu A_\mu
,\label{SUSYgauge}
\end{align}
and we consider the $(0,1)$-graded SUSY transformation together with a compensating gauge transformation,
\begin{align}
 \delta_{01}' V_{\rm WZ}:=(\delta_{01}+\delta_{\rm gauge})V_{\rm WZ},
\end{align}
where $\delta_{\rm gauge} V_{\rm WZ}:=i\big(\Lambda^{(1)}-\bar \Lambda^{(1)}\big)+ig\big[V_{\rm WZ},\Lambda^{(1)} +\bar\Lambda^{(1)}\big]$. Using this choice of $\Lambda^{(1)}$, the compensated transformation $\delta'_{01} V_{\rm WZ}$ preserves the Wess--Zumino gauge condition, and the explicit forms of component fields are given by
\begin{align}
    \delta_{01}' A_\mu
     & =i \varepsilon_{01} \sigma_\mu \bar \lambda_{01} +i \bar \varepsilon_{01} \bar \sigma_\mu \lambda_{01}
    ,\nn\\
    \delta_{01}'\left(\lambda_{01}\right)^{\ua}
     & =-\left(\varepsilon_{01} \sigma^{\mu\nu}\right)^{\ua} F_{\mu\nu} + i \left(\varepsilon_{01}\right)^{\ua} D ,
    \nn\\
    \delta_{01}' \left(\bar \lambda_{01}\right)^{\da}
     & =\left( \bar \sigma^{\mu\nu} \bar \varepsilon_{01} \right)^{\da} F_{\mu\nu} -i \left(\bar \varepsilon_{01}\right)^{\da} D,
    \nn\\
    \delta_{01}' D
     & = -\varepsilon_{01}\sigma^\mu D_\mu \bar \lambda_{01} - D_\mu \lambda_{01} \sigma^\mu \bar \varepsilon_{01}, \label{01SUSY}
\end{align}
where $D_\mu := \partial_\mu +ig\cdot  {\rm ad} A_\mu$ and $ig\cdot {\rm ad} F_{\mu\nu} :=[D_\mu, D_\nu]$.

Finally, we introduce the $(0,1)$-graded chiral spinor superfield
$(W_{01})_{\alpha}$.
Using $V_{\rm WZ}$ with the coordinate $y^\mu$:
\begin{align}
  V_{\text{WZ}} (y,\xi,\bar{\xi})
      = -\left(\xi \sigma^\mu \bar \xi \right)A_\mu\nn
     + i\xi\xi \bar \xi \bar \lambda_{01}
    - i\bar \xi \bar \xi \xi  \lambda_{01}
    +\frac{1}{2}\xi\xi \bar\xi\bar\xi (D-i\partial_\mu A^\mu),
\end{align}
we define the chiral spinor superﬁeld \cite{WessBagger} 
\begin{align}
\left(W_{01}\right)_\alpha(y,\xi) &:= -\frac{1}{8g} (\bar{D^{01}} \bar{D}^{01}) e^{-2 g V_{\rm WZ}} D^{01}_\alpha e^{2 g V_{\rm WZ}}.
\end{align}
The complex conjugate defines the anti-chiral superfield $\left(\bar W_{01}\right)_{\da}(\bar y,\bar\xi). $
One can verify that this superfield satisfies the chiral constraints:
\begin{align*}
 \bar{D}^{01}_{\da} \left(W_{01}\right)_\beta=D^{01}_\alpha \left(\bar{W}_{01}\right)_{\db}= 0,
\end{align*}
and is covariant under the gauge transformation:
$\left(W_{01}\right)_\alpha  \to  e^{-2i g\Lambda} {(W_{01})_\alpha }\, e^{2i g\Lambda}$.
The gauge-invariant field strength is encoded in the chiral spinor superfield. The component expression is given by
\begin{align}
\left(W_{01}\right)_{\ua}(y, \xi)
&=  -i \left(\lambda_{01}\right)_{\ua} +\xi_{\ua}D -i (\sigma^{\mu \nu} \xi)_{\ua} F_{\mu \nu}+\xi \xi \sigma^\mu_{\ua \db} {D}_\mu \bar{\lambda}_{01}^{\db},
\end{align}
 From the component expansion of $(W_{01})_{\alpha}$, one identifies the physical content of the vector multiplet: $A_\mu$ is the $(0,0)$-graded gauge field, $(\lambda_{01})_{\alpha}$ is its $(0,1)$-graded gaugino, and $D$ is a real auxiliary field. The auxiliary field $D$ does not propagate and becomes algebraically determined once the action is specified.

At this stage, we have introduced all dynamical component fields belonging to the minimal $\Z_2^2$-vector multiplet. In the next subsection, we construct the full version of the minimal $\Z_2^2$-superfield on the $\Z_2^2$-superspace $\mathfrak S$.

\subsection{The $\mathbb Z_2^2$-superfield }
We now construct the full version of the minimal $\Z_2^2$-vector multiplet on the complete $\Z_2^2$-superspace.
For simplicity, we define the coordinate $\tilde y= y+i(\eta \sigma^\mu \bar \eta)$ and the superfield by
\begin{align}
    \mathcal A_{11}(\tilde y, \xi, \eta)
    &=\Phi_{11}(y,\xi)|_{y=\tilde y}+\sqrt{2}\eta W_{01}(y,\xi)|_{y=\tilde y} -\eta\eta G_{11}(y,\xi)|_{y=\tilde y}, \label{FullSF}
\end{align}
where $G_{11}(\tilde y,\xi)$ is another $(1,1)$-graded chiral superfield on $(\tilde y,\xi)$. In this coordinate, the $(1,0)$-graded supercharges and the covariant derivatives can be written by
\begin{align}
    Q_{\ua}^{10}
     & =\partial_{\eta^\ua}
    , \quad
    \bar Q_{\da}^{10}
    =\partial_{\bar{\eta}^{\da}}-2i\left(\eta \sigma^\mu\right)_{\da} \partial_\mu
,\label{op:10susy_y}\\
    D_{\ua}^{10}
     & =\partial_{\eta^\ua}+2i\left(\sigma^\mu \bar \eta\right)_{\ua} \partial_\mu
    , \quad
    \bar D_{\da}^{10}
    =\partial_{\bar{\eta}^{\da}}
.\label{10cov.der_y}
\end{align}
By construction, $\mathcal A_{11}$ satisfies
\begin{align*}
 \bar D^{10}\mathcal A_{11}
=
\bar D^{01}\mathcal A_{11}
=
0.
\end{align*}
Using the SUSY transformation \eqref{tf:field}, we obtain $(1,0)$-graded SUSY transformations as
\begin{align}
    \delta_{10}\Phi_{11}
     & =\sqrt{2}\varepsilon_{10} W _{01},\nn
    \\
    \delta_{10} \left(W _{01}\right)_{\ua}
     & =-\sqrt{2}\left(\varepsilon_{10}\right)_{\ua}G_{11}
    +\sqrt{2}i\left(\sigma^{\mu} \bar \varepsilon_{10} \right)_{\ua}\partial_\mu\Phi_{11},\nn
    \\
    \delta_{10}G_{11}
     & = -\sqrt{2}i \left(\partial_\mu  W _{01} \sigma^\mu \bar \varepsilon_{10}\right). \label{10SUSY}
\end{align}

Now we have to point out the following:
\begin{enumerate}
\item
In this setting, the chiral superfield $\Phi_{11}$ becomes $\mathfrak{su}(N)$-valued, $\Phi_{11}=\sum_a \Phi_{11}^a T^a$, $T^a \in \mathfrak{su}(N)$,
since it belongs to the same multiplet as $(W_{01})_{\alpha}$ by Section~\ref{sec.2.2}. For the same reason, the gauge transformation of $\Phi_{11}$ takes the form
\[
\Phi_{11} \;\to\; e^{-2i g\Lambda}\, \Phi_{11}\, e^{2i g\Lambda}.
\]
Moreover, the $(0,1)$-graded SUSY transformation $\delta_{01}\Phi_{11}$ in \eqref{Phi01SUSY1} must be replaced by the gauge-covariant
transformation $\delta'_{01}$ defined in \eqref{SUSYgauge} with the coordinate $y^\mu$. The component expressions are
\begin{align}
    \delta'_{01}\varphi_{11}(y)
     & =\sqrt{2}\varepsilon_{01}\psi_{10},\nn
    \\
    \delta'_{01}\left(\psi_{10}(y)\right)_{\ua}
     & =-\sqrt{2}\left(\varepsilon_{01}\right)_{\ua}F_{11}
    +\sqrt{2}i\left(\sigma^{\mu} \bar \varepsilon_{01} \right)_{\ua}D_\mu\varphi_{11},\nn
    \\
    \delta'_{01}F_{11}(y)
     & = -\sqrt{2}i \left(D_\mu \psi_{10} \sigma^\mu \bar \varepsilon_{01}\right)
     -2ig[\bar\lambda_{01}\bar\varepsilon_{01},\varphi_{11}].
\label{Phi01SUSY2}
\end{align} 

\item
By Section~\ref{sec.2.2}, the vector multiplet consists of $\langle \varphi_{11},(\psi_{10})_{\alpha}\rangle \in \Phi_{11}$ and $\langle (\lambda_{01})_{\alpha},A_\mu\rangle \in (W_{01})_{\alpha}$. This shows that $G_{11}$
does not correspond to an independent physical degree of freedom.
To eliminate this redundancy, we determine $G_{11}$ uniquely
by imposing chirality, mass dimension, and gauge covariance. Namely, we require
\[
\bar D^{01}_{\dot\alpha}G_{11}=0,
\qquad
G_{11}\to e^{-2i g\Lambda} G_{11} e^{2i g\Lambda}
,
\]
and $[G_{11}]=2$.
\footnote{
We set the mass dimension as follows:
\begin{align*}
 [x]=-1,\quad [\xi]=[\eta]=-\frac12,\quad [\partial_\mu]=[A_\mu]=1,\quad [D^{01}]=[D^{10}]=\frac12.
\end{align*}
From this setting, we obtain
\begin{align*}
 [V_{\rm WZ}]=0,\quad [\Phi_{11}]=1,\quad [(W_{01})_{\ua}]=\frac32,\quad [G_{11}]=2.
\end{align*}
}
Under these conditions, $G_{11}$ is fixed as
\begin{align}
 G_{11}(y,\xi)
 =\frac14
 \bar D^{01}\bar D^{01}
 e^{-2gV_{\rm WZ}}
 \bar \Phi
 e^{2gV_{\rm WZ}} 
. \label{fixed_G}
\end{align}
The component expression is given by
\begin{align}
 G_{11}(y,\xi)
 &=\bar F_{11} 
+\xi\left(
\sqrt{2}i\sigma^\mu D_\mu \bar\psi_{10}
-2ig\{\lambda_{01},\bar\varphi_{11}\}
\right)
\nn\\
&+\xi\xi\left(
-D_\mu D^\mu \bar \varphi_{11}
+g[D,\bar\varphi_{11}]
+\sqrt{2}ig[\bar\lambda_{01},\bar\psi_{10}]
\right),
\end{align}
where $[\bar \lambda_{01},\bar\psi_{10}]\equiv [(\bar\lambda_{01})_{\da},\bar\psi_{10}^{\da}]$.
\item
The $(1,0)$-graded SUSY transformation \eqref{10SUSY} is not gauge covariant. Moreover, once $G_{11}$ is fixed by the defining relation above, the naive transformation $\delta_{10}$ does not preserve this constraint. 
In other words, the bare $(1,0)$-SUSY transformation is inconsistent with the gauge fixing of $G_{11}$.
Therefore, it is necessary to introduce a compensating gauge transformation
such that the combined transformation
\[
\delta'_{10}\mathcal A_{11}
=
(\delta_{10}+\delta_{\rm gauge})\mathcal A_{11}
\]
is gauge covariant and simultaneously preserves the defining relation of $G_{11}$.
The compensating transformation is taken as
\begin{align}
 \delta_{\rm gauge}\mathcal A_{11}
=
-2ig\big[\Lambda^{(2)},\mathcal A_{11}\big],
\label{01gauge}
\end{align}
where the explicit form of $\Lambda^{(2)}$ is not required,
since it will not be used in the subsequent analysis.
\end{enumerate}

We have thus completed the construction of the $\Z_2^2$-superfield formulation
on $\mathfrak S$, which will be used to build the $\Z_2^2$-SUSY invariant action. This formulation is consistent with both the $\Z_2^2$-superfield formulation in $(0+1)$ dimensions and the ordinary $\mathcal N=2$ superfield formulation.

\section{$\mathbb Z_2^2$-Graded supersymmetric Yang--Mills theory}
\subsection{The $\mathbb Z_2^2$-supersymmetric action}
The structure of the action is inspired by the holomorphic formulation of ordinary $\mathcal N=2$ SUSY Yang–Mills theory, appropriately generalized to the $\mathbb{Z}_2^2$-graded setting.
We define the $\mathbb Z_2^2$-SUSY invariant action by
\begin{align}
    S
    =\frac{1}{8\pi C(G)}\mathrm{Im}\left(
    \int \dd^4 \tilde y\, \dd^2\xi\, \dd^2\eta\,
    \tau\,\mathrm{Tr}\,\bigl(\mathcal A_{11}(\tilde y,\xi,\eta)\bigr)^2
    \right),
    \label{action}
\end{align}
where $C(G)$ denotes the quadratic Casimir in the adjoint representation of $\mathfrak{su}(N)$  defined by
\begin{align}
C(G) \delta^{ab} = \Tr(T^a T^b).
\end{align}
The holomorphic coupling constant is
\begin{align}
\tau = \frac{\Theta}{2 \pi} + \frac{4 \pi i }{g^2},
\end{align}
with $\Theta$ being the Yang--Mills angle.

The integration over the $\mathbb Z_2^2$-superspace is studied in \cite{Pon2, Pon1, NARI}. In this case, it is defined in analogy with the ordinary superspace formulation. 
The integrations over the $(0,1)$- and $(1,0)$-graded Grassmann coordinates are realized by differentiation. 
We adopt the normalization
\begin{align}
 \int \dd^2\xi\,(\xi\xi)=1,
 \qquad 
 \int \dd^2\eta\,(\eta\eta)=1.
\end{align}

Assuming that all fields have compact support, the integration over $\tilde y$ reduces to the standard spacetime integral,
\begin{align}
 \int \dd^4 \tilde y\, \dd^2 \xi\, \dd^2\eta ~ F(\tilde y, \xi, \eta)
 =
 \int \dd^4 x\, \dd^2 \xi\, \dd^2\eta~
 F(\tilde y, \xi, \eta)\big|_{\tilde y=x},
\end{align}
since the total derivative terms vanish.

Therefore, the action \eqref{action} becomes
\begin{align}
    S
    &=\frac{1}{8\pi C(G)}\mathrm{Im}\left(
    \int \dd^4 x\, \dd^2\xi\, \dd^2\eta\,
    \tau\,\mathrm{Tr}\,\bigl(\mathcal A_{11}\bigr)^2
    \right)
    \nonumber\\
    &=\frac{1}{8\pi C(G)}\mathrm{Im}\left(
    \int \dd^4 x\, \dd^2\xi\,
    \tau\,\mathrm{Tr}\,\bigl(W_{01}W_{01}-2G_{11}\Phi_{11}\bigr)
    \right)
    \nonumber\\
    &=\int \dd ^4x\, \left( L_{gauge} + L_{matter}\right).
    \label{action2}
\end{align}

The gauge part of the Lagrangian is
\begin{align}
L_{gauge} 
&= \frac{1}{g^2 C(G)} 
\Tr \left(
- \frac{1}{4}F_{\mu\nu}F^{\mu\nu}
-i \lambda_{01} \sigma^\mu D_\mu \bar{\lambda}_{01}
+\frac{1}{2}D^2
\right)
\nonumber\\
&\quad
+ \frac{\Theta}{32 \pi^2 C(G)} 
\Tr (F_{\mu\nu} \tilde{F}^{\mu \nu}),
\label{Lgauge}
\end{align}
while the matter part reads
\begin{align}
L_{matter}
&=\frac{1}{g^2 C(G)}
\Tr 
\Big(
-D_\mu \varphi_{11} D^\mu \bar \varphi_{11}
-i\psi_{10} \sigma^\mu D_\mu \bar \psi_{10}
+\bar F_{11} F_{11}
+gD[\varphi_{11},\bar\varphi_{11}]
\nonumber\\
&\qquad
+\sqrt{2}ig\big(
 \lambda_{01} \{\psi_{10},\bar \varphi_{11}\}
-\bar\psi_{10}\{\bar \lambda_{01}, \varphi_{11}\}
\big)
\Big),
\label{Lmatter}
\end{align}
where
\begin{align}
\tilde{F}^{\mu \nu}
= \frac{1}{2}\varepsilon^{\mu \nu \rho \sigma}F_{\rho \sigma},\quad \text{with}\quad \varepsilon^{0123}=-1,
\end{align}
which satisfies the Bianchi identity $D_\mu \tilde F^{\mu\nu}=0$.

Now we eliminate the non-propagating auxiliary fields $D, F, \bar F$. From their equations of motion, we obtain
\begin{align}
    D=-g[\varphi_{11},\bar\varphi_{11}],\quad F_{11}=\bar F_{11}=0.
\end{align}
Then we obtain the Lagrangian $\mathcal L= \mathcal L_{\rm gauge} +\mathcal L_{\rm matte}$ with
\begin{align}
    \mathcal L_{gauge}
    &= \frac{1}{g^2 C(G)} \Tr \left(- \frac{1}{4}F_{\mu\nu}F^{\mu\nu}-i \lambda_{01} \sigma^\mu D_\mu \bar{\lambda}_{01} \right)
+ \frac{\Theta}{32 \pi^2 C(G)} \Tr (F_{\mu\nu} \tilde{F}^{\mu \nu})
    ,\nn\\
    \mathcal L_{matter}
    &=\frac{1}{g^2 C(G)}\mathrm{Tr}\ 
    \bigg( -D_\mu \varphi_{11} D^\mu \bar \varphi_{11}
    -i\psi_{10} \sigma^\mu D_\mu \bar \psi_{10}
    -\frac{g^2}{2}\Big([\varphi_{11},\bar\varphi_{11}]\Big)^2
    \nonumber\\
    &\qquad\qquad\qquad\qquad\qquad 
    +\sqrt{2}ig\Big(
 \lambda_{01} \{\psi_{10},\bar \varphi_{11}\}
-\bar\psi_{10}\{\bar \lambda_{01}, \varphi_{11}\}
\Big)
    \bigg)
    .\label{Lag_on}
\end{align}

\begin{remark}
The action is manifestly $\mathbb{Z}_2^2$-graded supersymmetric, since both the superspace measure in \eqref{action} and the superfield $\mathcal A_{11}$ are invariant. 
Nevertheless, one may verify the invariance explicitly. 
Using the $(1,0)$-graded SUSY transformations \eqref{10SUSY}, together with the second line of \eqref{action2}, one confirms the $(1,0)$-SUSY invariance. 
Similarly, using the $(0,1)$-graded SUSY transformations \eqref{01SUSY} and \eqref{Phi01SUSY2}, along with the explicit form \eqref{Lag_on}, one verifies the $(0,1)$-SUSY invariance. 
Therefore, the action \eqref{action} is invariant under the $\mathbb{Z}_2^2$-graded supersymmetry.
Moreover, the resulting on-shell Lagrangian is invariant under the $R$-symmetry transformations \eqref{R00sym} and \eqref{R11sym}, provided that $r=0$ in \eqref{R00sym}.
\end{remark}
\medskip
\begin{remark}
The resulting action has the same structural form as conventional SUSY Yang--Mills theory. 
In particular, all kinetic terms appear with the standard sign, indicating the absence of classical instabilities associated with postive-sign kinetic terms. 
The interaction terms involve $\Z_2^2$-graded commutator and anticommutator structures, and thus differ from those of ordinary SUSY Yang--Mills theory. 
Furthermore, despite the presence of fields with generalized $\mathbb{Z}_2^2$-graded exchange properties, the positivity of the Hamiltonian follows from the $\Z_2^2$-SUSY algebra and is preserved by the structure of the action at the classical level. 
This suggests that the model remains stable even in the presence of generalized statistics.
\end{remark}

\subsection{The equations of motion and Noether currents}

We now derive the equations of motion.
For simplicity, we set the Yang--Mills angle $\Theta=0$.
From the Lagrangian \eqref{Lag_on}, we obtain
\begin{align}
 D_\mu D^\mu \varphi_{11} &=
 g^2 \left[\left[\varphi_{11},\bar \varphi_{11}\right],\varphi_{11}\right]
 -\sqrt{2}ig \left[\lambda_{01}, \psi_{10}\right],
 \nn\\
 D_\mu D^\mu \bar \varphi_{11} &=
 g^2 \left[\bar \varphi_{11},\left[\varphi_{11},\bar \varphi_{11}\right]\right]
 +\sqrt{2}ig \left[\bar \psi_{10}, \bar \lambda_{01}\right],
 \nn\\
 D_\mu F^{\mu\nu}&=
 g\Big(
 i\left[\bar\varphi_{11},D^\nu \varphi_{11}\right] 
 +i\left[\varphi_{11},D^\nu \bar \varphi_{11}\right]
 +\left\{\lambda_{01}\sigma^\nu, \bar \lambda_{01}\right\}
 +\left\{\psi_{10}\sigma^\nu, \bar\psi_{10}\right\}
 \Big),
 \nn\\
 \big(\bar\sigma^\mu D_\mu \lambda_{01}\big)^{\dot\alpha}&=
 -\sqrt{2} g \left\{\bar\psi_{10}^{\dot\alpha},\varphi_{11}\right\},
 \qquad
 \big(\sigma^\mu D_\mu \bar \lambda_{01}\big)_\alpha=
 \sqrt{2} g \left\{\psi_{10\,\alpha},\bar\varphi_{11}\right\},
 \nn\\
 \big(\bar\sigma^\mu D_\mu \psi_{10}\big)^{\dot\alpha}&=
 -\sqrt{2} g \left\{\bar\lambda_{01}^{\dot\alpha},\varphi_{11}\right\},
 \qquad
 \big(\sigma^\mu D_\mu \bar \psi_{10}\big)_\alpha=
 \sqrt{2} g \left\{\lambda_{01\,\alpha},\bar\varphi_{11}\right\}.
\label{EOM}
\end{align}

The structure of these equations is formally similar to that of ordinary $\mathcal N=2$ SUSY Yang--Mills theory. However, the interaction terms are governed by the $\mathbb Z_2^2$-graded commutation relations.
In particular, the Yukawa interactions and the scalar potential originate from the $\Z_2^2$-graded bracket structure. 

We now define the $(0,1)$-graded Noether current by
\begin{align}
    \frac{1}{g^2 C(G)}
    \varepsilon_{01}\partial_\mu J^\mu_{01} 
    :=
    \left.\left(\sum_{f}
    \partial_\mu\left((\delta_{01}' f)
    \frac{\partial \mathcal L}{\partial (\partial_\mu f)}\right)
-\delta_{01}' \mathcal L\right)\right|_{\varepsilon_{01}},
\end{align}
where 
\(
f\in 
\{A_\mu,\,
\varphi_{11},\,
\bar \varphi_{11},\,
\lambda_{01},\,
\bar \lambda_{01},\,
\psi_{10},\,
\bar \psi_{10}\}.
\)
By direct computation, we obtain
\begin{align}
(J_{01}^\mu)_\alpha
=\sqrt{2}(\sigma^\nu\bar\sigma^\mu\psi_{10})_{\alpha}
 D_\nu \bar\varphi_{11}
+g(\sigma^\mu\bar\lambda_{01})_{\alpha}
[\varphi_{11},\bar\varphi_{11}]
\nonumber\\
\qquad
-i(\sigma_\nu\bar \lambda_{01})_{\alpha}
F^{\mu\nu}
-(\sigma_\nu \bar \lambda_{01})_{\alpha}
\tilde F^{\mu\nu}
. \label{01current}
\end{align}
Using the equations of motion \eqref{EOM}, one verifies that the current is conserved on-shell,
\begin{align}
 {\rm Tr}\,\int \dd^4 x~ \partial_\mu (J_{01}^{\mu})_{\ua}
 ={\rm Tr}\,\int \dd^4 x~ D_\mu (J_{01}^{\mu})_{\ua}
=0.
\end{align}

We now define the $(1,0)$-graded Noether current through the $R^{11}$-transformation \eqref{R11sym}.
Since
\begin{align}
 \delta_{R^{11}} (J^\mu_{01})_{\alpha}
 \propto i\gamma_{11}(J^\mu_{10})_{\alpha},
\end{align}
this motivates the definition
\begin{align}
 (J_{10}^\mu)_\alpha
=\sqrt{2}
(\sigma^\nu\bar\sigma^\mu\lambda_{01})_{\alpha}
 D_\nu \bar\varphi_{11}
+g(\sigma^\mu\bar\psi_{10})_{\alpha}
[\varphi_{11},\bar\varphi_{11}]
\nonumber\\
\qquad
-i(\sigma_\nu\bar \psi_{10})_{\alpha}
F^{\mu\nu}
-(\sigma_\nu \bar \psi_{10})_{\alpha}
\tilde F^{\mu\nu}
.\label{10current}
\end{align}

Again, using \eqref{EOM}, one verifies that
\begin{align}
 {\rm Tr}\,\int \dd^4 x~ \partial_\mu (J_{10}^{\mu})_{\ua}
 ={\rm Tr}\,\int \dd^4 x~ D_\mu (J_{10}^{\mu})_{\ua}
=0
\end{align}
holds on-shell.

The $(1,0)$-graded current may not be obtained directly from the $(1,0)$-graded SUSY transformation \eqref{10SUSY}. However, this does not lead to any inconsistency, since the $(1,0)$-graded transformation contains a compensating gauge transformation. With an appropriate choice of gauge representative, the same current should be reproduced. 

\section{Concluding remarks}

In this paper, we have constructed a classical minimal $\mathbb{Z}_2^2$-SUSY Yang--Mills theory using a superfield formulation. 
We explicitly derived the invariant action \eqref{action}, the equations of motion~\eqref{EOM}, and the associated Noether currents~(\ref{01current},~\ref{10current}), thereby establishing the internal consistency of the model at the classical level.

The resulting action has the same structural form as conventional supersymmetric Yang--Mills theory. 
In particular, all kinetic terms in \eqref{Lag_on} appear with the standard sign, indicating the absence of classical ghost-like instabilities associated with wrong-sign kinetic terms. 
Moreover, the positivity of the Hamiltonian follows from the underlying supersymmetry algebra and is compatible with the structure of the action, even in the presence of fields with generalized $\mathbb{Z}_2^2$-graded exchange properties.

These results demonstrate that generalized statistics arising from $\mathbb{Z}_2^2$-grading can be realized at the classical level in a stable interacting supersymmetric gauge theory without introducing classical instabilities. 
In particular, this provides strong evidence that the conventional relation between spin, statistics, and stability admits a broader realization when extended through generalized grading structures.

Several important questions remain open. 
In particular, the quantization of the model, the structure of the Hilbert space, and the possible emergence of anomalies or quantum instabilities require further investigation. 
It would also be interesting to explore the implications for scattering amplitudes and the extension to other classes of gauge and supergravity theories.

\section*{Acknowledgements}
The authors would like to thank N. Aizawa , N. Maru, and T. Nishinaka for valuable discussions.  R. I., A. N., and S. T. are supported by JST SPRING, Grant Number JPMJSP2139.

\bibliographystyle{JHEP}
\bibliography{refs}

\end{document}